\documentclass[conference, a4paper]{IEEEtran}
\IEEEoverridecommandlockouts
\usepackage{cite}
\usepackage{amsmath,amssymb,amsfonts}
\usepackage{algorithmic}
\usepackage{graphicx}
\usepackage{textcomp}
\usepackage{xcolor}
\usepackage[left=1.57cm,right=1.57cm,top=0.95cm,bottom=2.54cm]{geometry}
\def\BibTeX{{\rm B\kern-.05em{\sc i\kern-.025em b}\kern-.08em
    T\kern-.1667em\lower.7ex\hbox{E}\kern-.125emX}}

\usepackage{hyperref}
\usepackage{booktabs}  
\usepackage{amssymb}   
\usepackage{multirow}
\usepackage{array}
\usepackage{tabularx}
\usepackage{booktabs}
\usepackage{amsmath,amssymb}
\usepackage{array,tabularx,booktabs}
\usepackage{pifont, amssymb}   
\usepackage{hyperref}
\usepackage{fontawesome5} 
\usepackage{xcolor}
\usepackage{tikz}
\usepackage{makecell}

\definecolor{myorange}{HTML}{FD742D}
\definecolor{mygreen}{HTML}{185A56}
 
\newcommand{\autonumlabel}[1]{\refstepcounter{equation}\label{#1}(\theequation)}
\hypersetup{
    colorlinks=true,
    linkcolor=orange,   
    citecolor=green!30!black,    
    urlcolor=orange      
}
\def\BibTeX{{\rm B\kern-.05em{\sc i\kern-.025em b}\kern-.08em
    T\kern-.1667em\lower.7ex\hbox{E}\kern-.125emX}}

\begin{document}

\title{Leveraging Optimal Information-Power Flow for Transmission Switching in AC/MTDC Grids\\
\thanks{This work was funded  by the CRESYM project Harmony (\href{URL}{https://cresym.eu/harmony/}).}
}

\author{\IEEEauthorblockN{1\textsuperscript{st} Haixiao Li}
\IEEEauthorblockA{\textit{Department of Electrical Sustainable Energy} \\
\textit{Delft University of Technology}\\
Delft, Netherlands \\
h.li-16@tudelft.nl}
\and
\IEEEauthorblockN{2\textsuperscript{nd} Aleksandra Leki\'{c}}
\IEEEauthorblockA{\textit{Department of Electrical Sustainable Energy} \\
\textit{Delft University of Technology}\\
Delft, Netherlands \\
a.lekic@tudelft.nl}

}

\IEEEoverridecommandlockouts
\IEEEpubid{\makebox[\columnwidth]{XXX-X-XXXX-XXXX-X/XX/\$XX.XX~\copyright20XX IEEE \hfill} \hspace{\columnsep}\makebox[\columnwidth]{ }}

\maketitle

\IEEEpubidadjcol

\begin{abstract}
The emerging AC/multi-terminal DC grids are regarded as a promising solution for accommodating the increasing integration of renewable energy sources. This work proposes an optimization framework to address transmission switching (TS) problems arising in practical operational scenarios, such as maintenance scheduling, contingency management, and fault restoration. Unlike most existing studies, the proposed framework considers the role of communication networks in TS operations and develops an optimal information-power flow (OIPF) model. The OIPF model captures the impact of information flows on circuit breaker actions while incorporating communication-related costs, thereby better reflecting practical operational decision-making processes. To ensure computational tractability, the resulting optimization problem is formulated as a mixed-integer second-order cone programming (MISOCP) model through convex relaxations, polygonal approximations, and Big-M reformulations. Numerical case studies illustrate the applicability of the proposed OIPF model and indicate its potential in supporting transmission switching decisions.
\end{abstract}


\section{Introduction}
\subsection{Background and Motivation}
Around the world, there is a strong desire to develop renewable energy source (RES) in order to alleviate the energy and environmental crises. Against this backdrop, an increasing number of offshore RES plants are being planned and gradually commissioned \cite{R1}. To deliver renewable power to onshore AC grids, voltage-source-converter (VSC)-based multi-terminal DC (MTDC) systems are emerging as a promising transmission infrastructure, enabling flexible power transfer \cite{R2}. In practical operation, such AC/MTDC grids may encounter various scenarios, including line maintenance outages, fault recovery, and other unforeseen contingencies. Under these conditions, the original transmission topology may need to be reconfigured into another feasible network structure. Consequently, there is a strong need to determine a rational and cost-effective topology reconfiguration scheme from a set of candidate network topologies.

\subsection{Related Works}
\textit{Transmission switching} (TS) problems, also referred to as topology reconfiguration problems, aim to determine the binary on/off states of transmission switches such that the power system satisfies physical operating constraints while achieving economical operation. The optimal TS problem with nonlinear power flow equations is inherently formulated as a mixed-integer nonlinear programming (MINLP) problem, which is generally computationally challenging to solve. To address this issue, alternative power flow formulations, such as linearized \cite{R10} or convexified models \cite{R11, R12}, are commonly adopted. Although these approximations inevitably introduce certain modeling inaccuracies, they significantly improve the computational tractability of the resulting optimization problem. Moreover, although the overall formulation remains nonconvex due to the presence of binary variables, the continuous relaxation of the model can still preserve convexity and thus be efficiently handled by off-the-shelf optimization solvers.

Following this line of research, a number of excellent works \cite{R3, R4, R5, R6, R7, R9} have been developed to address optimal TS problems in AC networks, DC networks, and both. These studies have significantly advanced the modeling and solution methodologies for topology reconfiguration problems. Nevertheless, one aspect has received comparatively limited attention, namely how the information required for switching actions is delivered and coordinated. This issue becomes particularly important in AC/MTDC grids, where power transmission often spans long geographical distances (e.g., hundreds to thousands of kilometers), and VSC stations are geographically dispersed. Under such conditions, switching actions in MTDC grids can no longer be regarded as purely local operations. Instead, they rely on coordinated information exchange among multiple remote terminals. Therefore, beyond optimizing power flow itself, it is also important to consider how switching-related information can be properly incorporated into the conventional \textit{optimal power flow} (OPF) paradigm.

\begin{table}
\caption{Comparison of representative TS studies.}
\label{tab:1}
\centering
\footnotesize
\setlength{\tabcolsep}{3.5pt}
\renewcommand{\arraystretch}{1.15}

\begin{tabular}{c c cc cc}
\toprule
\multirow{2}{*}{\textbf{References}} 
& \multirow{2}{*}{\textbf{Formulation}} 
& \multicolumn{2}{c}{\textbf{Transmission Switching}} 
& \multicolumn{2}{c}{\textbf{Flow}} \\

\cmidrule(lr){3-4} \cmidrule(lr){5-6}
& 
& AC Net. & DC Net. 
& Pow. & Info. \\
\midrule

\cite{R3} & MILP & $\checkmark$ & $\times$ & $\checkmark$ & $\times$ \\
\cite{R4} & MISOCP & $\checkmark$ & $\times$ & $\checkmark$ & $\times$ \\
\cite{R5} & MIQCP & $\checkmark$ & $\times$  & $\checkmark$ & $\times$ \\
\cite{R6} & MISOCP & $\times$ & $\checkmark$ & $\checkmark$ & $\times$ \\
\cite{R7} & MISOCP & $\times$ & $\checkmark$ & $\checkmark$ & $\times$ \\
\cite{R9} & MILP, MISOCP & $\checkmark$ & $\checkmark$ & $\checkmark$ & $\times$ \\
\midrule

\textbf{This Work} &  MISOCP &  $\times$ & $\checkmark$ & $\checkmark$ & $\checkmark$ \\

\bottomrule
\end{tabular}
\vspace{0.5pt}
\begin{flushleft}
\footnotesize
\ding{71} ``$\checkmark$'' means involved.  ``$\times$'' means not involved. ``MILP'' refers to mixed-integer linear programming. ``MIQCP'' refers to mixed-integer quadratically constrained programming. ``MISOCP'' refers to mixed-integer second-order cone programming. 
\end{flushleft}
\end{table}

\subsection{Contributions}
Building upon this perspective, this paper proposes an \textit{optimal information--power flow} (OIPF) model to support optimal TS in AC/MTDC grids. The proposed OIPF framework explicitly captures the interaction between power flow and information flow during TS operations. By incorporating information flow constraints and associated communication costs, the model attempts to reflect certain practical limitations and overhead of communication infrastructures involved in switching coordination. Although the consideration of information flow in this work is still at a preliminary stage, the proposed framework introduces a new perspective for investigating TS problems. Considering computational scalability, the resulting OIPF model is formulated as a \textit{mixed-integer second-order cone programming} (MISOCP) problem, facilitating its solution with reasonable computational effort. Table~\ref{tab:1} provides an overview of recent TS studies and highlights the main contributions of this work.

The rest of the paper is structured as follows: Section~\ref{section:2} formulates the OIPF model. Section~\ref{section:3} provides case studies to validate the proposed OIPF model, followed by conclusions drawn in Section~\ref{section:4}.

\section{Model Formulation}\label{section:2}

\subsection{Constraints for Power Flow in AC/MTDC Grids}
\textbf{\textit{Sets, indexes, and scripts}}:
$\mathcal{N}/\mathcal{E}$ denote the system-wide node/branch sets.
$(\bar{\bullet})/(\underline{\bullet})$ denote the upper/lower bounds of variables.
$({\bullet})^{ac}/({\bullet})^{dc}/({\bullet})^{vsc}/({\bullet})^{res}$ denote the parameters or variables associated with the AC grid/MTDC grid/VSC station/RES plant.

\textbf{\textit{Constraints of AC grids}}: The formulations are summarized in Table~\ref{tab:2}. \eqref{eq:1} formulates the AC branch flows. $p_{i j}^{a c} / q_{i j}^{a c}$ denotes the AC real/reactive branch power. $g_{i j}^{a c} / b_{i j}^{a c}$ denotes the AC branch conductance/susceptance. $c_{i i}^{a c}, c_{i j}^{a c}, s_{i j}^{a c}$ are the introduced axillary variables related to the square of AC bus voltage $v_{i}^{a c}$. More specifically, given that $e_{i}^{a c}:=\Re\{v_{i}^{a c}\}$ and $f_{i}^{a c}:=\Im\{v_{i}^{a c}\}$, we have that $c_{i j}^{ac}:=e_i^{ac}e_j^{ac}+f_i^{ac} f_j^{ac}$ and  $s_{i j}^{ac}:=e_i^{ac} f_j^{ac}-e_j^{ac} f_i^{ac}$; \eqref{eq:2} regulates the limit regarding AC branch capacity $\bar{s}_{i j}^{a c}$, using $N$-polygonal approximation \cite{R13} to approximate the original nonlinear expression that $\left(p_{ij}^{ac}\right)^2+\left(q_{ij}^{ac}\right)^2\leq\left(\bar{s}_{i j}^{a c}\right)^2$; \eqref{eq:3} formulates the AC bus power injections. $p_{i}^{a c} / q_{i}^{a c}$ denotes the AC real/reactive bus  power injection. $G_{i, s h}^{a c} / B_{i, s h}^{a c}$ denotes the real/imaginary parts of the shunt admittance of the AC buses; \eqref{eq:4} specifies the composition of $p_{i}^{a c}$ and $q_{i}^{a c}$. $p_{i, g e n}^{a c} / q_{i, g e n}^{a c}$ denotes the real/reactive generator power output. $p_{i, load}^{a c} / q_{i, load}^{a c}$ denotes the real/reactive load consumption. $p_{i, a2v}^{a c} / q_{i, a2v}^{a c}$ denotes the real/reactive power transmission, which is transmitted from the AC grid to the VSC station; \eqref{eq:5} formulates the bounds of the squared AC bus voltage, as $c_{ii}^{ac}$ is essentially equivalent to $\left(v_{i}^{ac}\right)^2$; \eqref{eq:6} regulates the bounds of generator outputs; \eqref{eq:7} forms an SOCP relaxation of the exact equation derived from AC power flow physics that $\left(c_{i j}^{a c}\right)^2+\left(s_{i j}^{a c}\right)^2 = c_{i i}^{a c} c_{j j}^{a c}$ \cite{R12}. 

\begin{table*}[t]
\centering
\caption{Constraints of AC Grids}
\label{tab:2}
\renewcommand{\arraystretch}{0.0}
\setlength{\tabcolsep}{4pt}

\begin{tabularx}{\textwidth}{@{}>{\raggedright\arraybackslash}X >{\centering\arraybackslash}p{1.2cm}@{}}
\toprule

\multicolumn{2}{@{}l@{}}{\textbf{AC branches}} \\

$\displaystyle
p_{i j}^{a c}=g_{i j}^{a c}\left(c_{i i}^{a c}-c_{i j}^{a c}\right)+b_{i j}^{a c} s_{i j}^{a c}, \quad
q_{i j}^{a c}=-b_{i j}^{a c}\left(c_{i i}^{a c}-c_{i j}^{a c}\right)+g_{i j}^{a c} s_{i j}^{a c}, \quad
\quad \forall(ij) \in \mathcal{E}^{a c}
$
& \autonumlabel{eq:1} \\ 

$\displaystyle
-\bar{s}_{ij}^{a c} \leq \cos \left(\frac{k}{N} \pi\right) p_{ij}^{a c}+\sin \left(\frac{k}{N} \pi\right) q_{ij}^{a c} \leq \bar{s}_{ij}^{a c},
\quad \forall k \in\{1,2, \cdots, N\}, \quad \forall(ij) \in \mathcal{E}^{a c}, 
$
& \autonumlabel{eq:2} \\ 

\addlinespace[0.2em]
\multicolumn{2}{@{}l@{}}{\textbf{AC buses}} \\

$\displaystyle
p_i^{a c}=\sum_{i j} p_{i j}^{a c}+c_{i i}^{a c} G_{i, s h}^{a c}, \quad q_i^{a c}=\sum_{i j} q_{i j}^{a c}-c_{i i}^{a c} B_{i, s h}^{a c},
\quad \forall i \in \mathcal{N}^{a c}, \quad \forall(ij) \in \mathcal{E}^{a c}
$
& \autonumlabel{eq:3} \\

$\displaystyle
p_i^{a c}=p_{i, g e n}^{a c}-p_{i, l o a d}^{a c}-p_{i, a 2 v}^{a c}, \quad q_i^{a c}=q_{i, g e n}^{a c}-q_{i, l o a d}^{a c}-q_{i, a 2 v}^{a c},
\quad \forall i \in \mathcal{N}^{a c}
$
& \autonumlabel{eq:4} \\

$\displaystyle
\left(\underline{v}_i^{a c}\right)^2 \leq c_{i i}^{a c} \leq\left(\bar{v}_i^{a c}\right)^2, \quad \forall i \in \mathcal{N}^{a c}
$
& \autonumlabel{eq:5} \\

\addlinespace[0.2em]
\multicolumn{2}{@{}l@{}}{\textbf{Generators}} \\

$\displaystyle
\underline{p}_{i, g e n}^{a c} \leq p_{i, g e n}^{a c} \leq \bar{p}_{i, g e n}^{a c}, \quad \underline{q}_{i, g e n}^{a c} \leq q_{i, g e n}^{a c} \leq \bar{q}_{i, g e n}^{a c},
\quad \forall i \in \mathcal{N}^{a c}
$
& \autonumlabel{eq:6} \\

\addlinespace[0.2em]
\multicolumn{2}{@{}l@{}}{\textbf{SOCP relaxation}} \\
$\displaystyle
\left(c_{i j}^{a c}\right)^2+\left(s_{i j}^{a c}\right)^2 \leq c_{i i}^{a c} c_{j j}^{a c}, \quad \forall i \in \mathcal{N}^{a c}, \quad \forall(ij) \in \mathcal{E}^{a c}
$
& \autonumlabel{eq:7} \\

\bottomrule
\end{tabularx}
\end{table*}

\textbf{\textit{Constraints of MTDC grids}}: The formulations are summarized in Table~\ref{tab:3}. \eqref{eq:8} formulates the DC branch power loss equation. $p_{j h}^{d c}$ denotes the DC branch power. $r_{j h}^{d c}$ denotes the DC branch resistance. $l_{j h}^{d c}$ denotes the squared DC branch current; \eqref{eq:9} formulates the DC branch voltage drop, where $u_j^{dc}$ denotes the square of DC bus voltage $v_j^{dc}$; \eqref{eq:10} regulates the DC branch power limit; \eqref{eq:11} formulates the DC bus power injections. $\rho^{d c}$ represents the polarity of the MTDC grid; Specifically, $\rho^{dc}:=1$ corresponds to a monopolar configuration, while $\rho^{dc}:=2$ corresponds to a bipolar configuration;  \eqref{eq:12} regulates the bounds of the squared DC voltage; \eqref{eq:13} forms an SOCP relaxation of the exact equation derived from DC power flow physics that $\left(p_{j h}^{d c}\right)^2 = l_{j h}^{d c} u_j^{d c}$ \cite{R4}. 

\begin{table*}[t]
\centering
\caption{Constraints of MTDC Grids}
\label{tab:3}
\renewcommand{\arraystretch}{0.0}
\setlength{\tabcolsep}{4pt}

\begin{tabularx}{\textwidth}{@{}>{\raggedright\arraybackslash}X >{\centering\arraybackslash}p{1.2cm}@{}}
\toprule

\multicolumn{2}{@{}l@{}}{\textbf{DC branches}} \\

$\displaystyle
p_{j h}^{d c}+p_{h j}^{d c}=r_{j h}^{d c} l_{j h}^{d c}, \quad \forall(jh),(hj) \in \mathcal{E}^{d c}
$
& \autonumlabel{eq:8} \\ 

$\displaystyle
u_j^{d c}-u_h^{d c}=r_{j h}^{d c}\left(p_{j h}^{d c}-p_{h j}^{d c}\right), \quad \forall j,h \in \mathcal{N}^{d c}, \quad \forall(jh),(hj) \in \mathcal{E}^{d c}
$
& \autonumlabel{eq:9} \\ 

$\displaystyle
\underline{p}_{jh}^{d c} \leq p_{jh}^{d c} \leq\bar{p}_{jh}^{d c}, \quad \forall(jh) \in \mathcal{E}^{d c}
$
& \autonumlabel{eq:10} \\

\addlinespace[0.2em]
\multicolumn{2}{@{}l@{}}{\textbf{DC buses}} \\

$\displaystyle
p_j^{d c}=\rho^{d c} \sum_{j h} p_{j h}^{d c}, \quad \forall j \in \mathcal{N}^{d c}, \quad \forall(jh) \in \mathcal{E}^{d c}
$
& \autonumlabel{eq:11} \\

$\displaystyle
\left(\underline{v}_j^{d c}\right)^2 \leq u_j^{d c} \leq\left(\bar{v}_j^{d c}\right)^2, \quad \forall j \in \mathcal{N}^{d c}
$
& \autonumlabel{eq:12} \\

\addlinespace[0.2em]
\multicolumn{2}{@{}l@{}}{\textbf{SOCP relaxation}} \\
$\displaystyle
\left(p_{j h}^{d c}\right)^2 \leq l_{j h}^{d c} u_j^{d c}, \quad \forall j \in \mathcal{N}^{d c}, \quad \forall(jh) \in \mathcal{E}^{d c}
$
& \autonumlabel{eq:13} \\

\bottomrule
\end{tabularx}
\end{table*}

\begin{figure}
    \centering
    \includegraphics[width=0.9\linewidth]{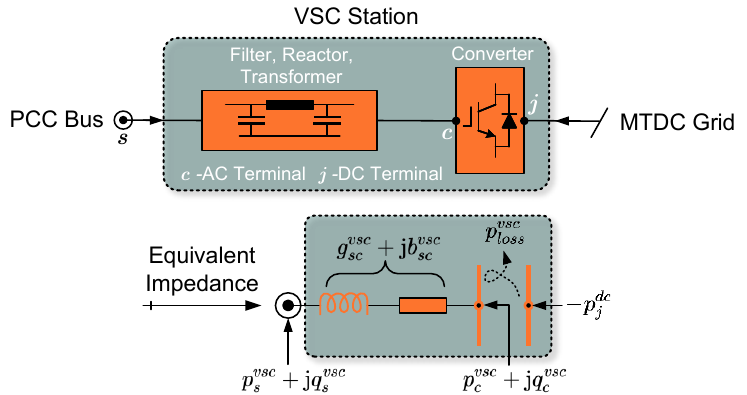}
    \caption{Equivalent impedance model of the VSC station. Considering VSC adopts modular multilevel topology, the equivalent filter capacitance can be ignored and not included in the equivalent impedance model.}
    \label{fig:1}
\end{figure}

\textbf{\textit{Constraints of RES plants}}: The formulations are summarized in Table~\ref{tab:5}. The RES plant is treated as an integrated node, thus the detailed internal power flow relationships are not involved. \eqref{eq:14} indicates that the RES power outputs $p_{out }^{res}, q_{out }^{res}$ exact to the transmitted power $p_{r2v }^{res}, q_{r2v }^{res}$, which are from the RES plant to the VSC station; \eqref{eq:15} regulates the bounds of RES power outputs. It involves two aspects: First, $p_{out }^{res}$ is subject to an upper bound imposed by natural factors. Second, $p_{out }^{res}$ and $q_{out }^{res}$ are subject to the capacity constraint of the RES station, which is formulated using a set of linearized capacity constraints similar to \eqref{eq:2}; \eqref{eq:33} regulates the bounds of the squared RES voltage at the point of common coupling (PCC), where $u_{pcc}^{res}$ denotes the square of RES PCC voltage $v_{pcc}^{res}$. 

\begin{table*}[t]
\centering
\caption{Constraints of RES Plants}
\label{tab:5}
\renewcommand{\arraystretch}{0.0}
\setlength{\tabcolsep}{4pt}

\begin{tabularx}{\textwidth}{@{}>{\raggedright\arraybackslash}X >{\centering\arraybackslash}p{1.2cm}@{}}
\toprule

\multicolumn{2}{@{}l@{}}{{\textbf{RES power}}} \\

$\displaystyle
p_{out}^{res}=p_{r2v}^{res}, \quad q_{out}^{res}=q_{r2v}^{res},
$
& \autonumlabel{eq:14} \\ 

$\displaystyle
0 \leq p_{out}^{res} \leq \bar{p}_{out}^{res}, \quad -\bar{s}_{o u t}^{res} \leq \cos \left(\frac{k}{N} \pi\right) p_{out}^{res}+\sin \left(\frac{k}{N} \pi\right) q_{out}^{res} \leq \bar{s}_{out}^{res},
\quad \forall k \in\{1,2, \cdots, N\}$ 
& \autonumlabel{eq:15} \\ 

\multicolumn{2}{@{}l@{}}{{\textbf{RES voltage}}} \\

$\displaystyle
\left(\underline{v}_{p c c}^{r e s}\right)^2 \leq u_{p c c}^{r e s} \leq\left(\bar{v}_{p c c}^{r e s}\right)^2, \quad \forall j \in \mathcal{N}^{r e s}
$
& \autonumlabel{eq:33} \\

\bottomrule
\end{tabularx}
\end{table*}

\textbf{\textit{Constraints of VSC stations}}: The formulations are summarized in Table~\ref{tab:4}. \eqref{eq:16} and \eqref{eq:17} formulate the VSC bus power injections. As illustrated in Fig.~\ref{fig:1}, $p_s^{v sc}/q_s^{vsc}$ denotes the real/reactive power injection at the PCC and $p_c^{v sc}/q_c^{vsc}$ denotes the real/reactive power injection at the AC terminal. $g_{s c}^{vsc} / b_{s c}^{vsc}$ denotes the branch conductance/susceptance from the PCC bus to the AC terminal. Similar to $c_{i i}^{a c}, c_{i j}^{a c}, s_{i j}^{a c}$ in \eqref{eq:1}, $c_{ss }^{vsc}, c_{s c}^{vsc}, s_{s c}^{vsc}, c_{cc}^{vsc}, c_{c s}^{vsc}, s_{c s}^{vsc}$ are the terms associated with the VSC bus voltage; \eqref{eq:18} specifies the bus power injection and voltage at PCC, which may be coupled with either $p_{i,a2v}^{ac}$, $q_{i,a2v}^{ac}$, and $c_{ii}^{ac}$, or $p_{r2v}^{res}$, $q_{r2v}^{res}$, and $u_{pcc}^{res}$, depending on whether the VSC is connected to the AC grid or the RES plant; \eqref{eq:19} denotes the VSC power loss $p_{loss }^{v s c}$, and as formulated in \eqref{eq:20}, $p_{loss }^{v s c}$ is subject to the quadratic function of the VSC current amplitude $I_c^{vsc}$. $l_c^{vsc}$ is used to denote the squared VSC current. $a_{loss}^{v s c}$, $b_{loss}^{v s c}$, $c_{loss}^{v s c}$ are VSC loss coefficients; \eqref{eq:21} regulates the bounds of $I_c^{vsc}$ and $l_c^{vsc}$; \eqref{eq:22} forms a SOCP relaxation of the original relationships that $\left(I_c^{v s c}\right)^2 = l_c^{v s c}$ and $\left(p_c^{v s c}\right)^2+\left(q_c^{v s c}\right)^2 = c_{c c}^{v s c} l_c^{v s c}$.

\begin{table*}[t]
\centering
\caption{Constraints of VSC Stations}
\label{tab:4}
\renewcommand{\arraystretch}{0.0}
\setlength{\tabcolsep}{4pt}

\begin{tabularx}{\textwidth}{@{}>{\raggedright\arraybackslash}X >{\centering\arraybackslash}p{1.2cm}@{}}
\toprule

\multicolumn{2}{@{}l@{}}{{\textbf{VSC Buses}}} \\

$\displaystyle
p_{s}^{v s c}=c_{s s}^{v s c} g_{s c}^{v s c}-c_{s c}^{v s c} g_{s c}^{v s c}+s_{s c}^{v s c} b_{s c}^{v s c}, \quad q_{s}^{v s c}=-c_{s s}^{v s c} b_{s c}^{v s c}+c_{s c}^{v s c} b_{s c}^{v s c}+s_{s c}^{v s c} g_{s c}^{v s c}, \quad s,c \in \mathcal{N}^{v s c}
$
& \autonumlabel{eq:16} \\ 

$\displaystyle
p_{c}^{v s c}=c_{c c}^{v s c} g_{c s}^{v s c}-c_{c s}^{v s c} g_{c s}^{v s c}+s_{c s}^{v s c} b_{c s}^{v s c}, \quad q_{c}^{v s c}=-c_{c c}^{v s c} g_{c s}^{v s c}+c_{c s}^{v s c} b_{c s}^{v s c}+s_{c s}^{v s c} g_{c s}^{v s c}, \quad c,s \in \mathcal{N}^{v s c}
$
& \autonumlabel{eq:17} \\ 

$\displaystyle
p_{s}^{vsc}=
\begin{cases}
p_{i,a2v}^{ac}, & \text{if }s \text{ is connected to the AC grid} \\
p_{r2v}^{res},& \text{if }s \text{ is connected to the RES plant} 
\end{cases},
\quad
q_{s}^{vsc}=
\begin{cases}
q_{i,a2v}^{ac}, & \text{if }s \text{ is connected to the AC grid} \\
q_{r2v}^{res}, & \text{if }s \text{ is connected to the RES plant} 
\end{cases}, \quad s \in \mathcal{N}^{v s c}, \quad i \in \mathcal{N}^{a c}
$
& \autonumlabel{eq:18} \\ 

\addlinespace[0.2em]
\multicolumn{2}{@{}l@{}}{{\textbf{VSC Power Loss}}} \\

$\displaystyle
p_c^{v s c}+p_{loss}^{v s c}+p_j^{d c}=0,  \quad c \in \mathcal{N}^{v s c}, \quad j \in \mathcal{N}^{d c}
$
& \autonumlabel{eq:19} \\ 

$\displaystyle
p_{loss}^{v s c}=a_{loss}^{v s c}+b_{loss}^{v s c} I_c^{v s c}+c_{loss}^{v s c} l_c^{v s c}, \quad c \in \mathcal{N}^{v s c}
$
& \autonumlabel{eq:20} \\ 

$\displaystyle
0 \leq I_c^{v s c} \leq \bar{I}_c^{v s c}, \quad 0 \leq l_c^{v s c} \leq (\bar{I}_c^{v s c})^2, \quad c \in \mathcal{N}^{v s c}
$
& \autonumlabel{eq:21} \\ 

\addlinespace[0.2em]
\multicolumn{2}{@{}l@{}}{{\textbf{SOCP Relaxation}}} \\
$\displaystyle
\left(I_c^{v s c}\right)^2 \leq l_c^{v s c}, \quad \left(p_c^{v s c}\right)^2+\left(q_c^{v s c}\right)^2 \leq c_{c c}^{v s c} l_c^{v s c}, \quad c \in \mathcal{N}^{v s c}
$
& \autonumlabel{eq:22} \\

\bottomrule
\end{tabularx}
\end{table*}

\subsection{Constraints for DC Transmission Switching}
In this work, we especially consider TS for the MTDC grid. As presented in \cite{R6, R7}, binary variables $\alpha_{jh}^{dc}$ are introduced to represent the connection status of DC branches. $\alpha_{jh}^{dc}:=1$ indicates that the DC branch $jh$ is connected, and otherwise is disconnected. Particularly, $\alpha_{jh}^{dc}$ has the inherent symmetry that $\alpha_{jh}^{dc}=\alpha_{hj}^{dc}$. With the introduced $\alpha_{jh}^{dc}$, \eqref{eq:9} is no longer applicable and needs to be modified, resulting in the following constraints, such that \cite{R7}:
\begin{subequations}\label{eq:23}
\begin{equation}\label{eq:23a}
- \mathrm{M} \alpha_{jh}^{dc} \leq p_{jh}^{dc} \leq \mathrm{M} \alpha_{jh}^{dc},
\end{equation}
\begin{equation}\label{eq:23b}
(u_j^{dc}-b_{jh}^{dc})-(u_h^{dc}-t_{jh}^{dc}) = r_{jh}^{dc}(p_{jh}^{dc}-p_{hj}^{dc}),
\end{equation}
\begin{equation}\label{eq:23c}
\left(\underline{v}_j^{dc}\right)^2(1-\alpha_{jh}^{dc}) \leq b_{jh}^{dc} \leq \left(\bar{v}_j^{dc}\right)^2(1-\alpha_{jh}^{dc}),
\end{equation}
\begin{equation}\label{eq:23d}
\left(\underline{v}_h^{dc}\right)^2(1-\alpha_{jh}^{dc}) \leq t_{jh}^{dc} \leq \left(\bar{v}_h^{dc}\right)^2(1-\alpha_{jh}^{dc}),
\end{equation}
\begin{equation}\label{eq:23e}
\left(\underline{v}_j^{dc}\right)^2 \alpha_{jh}^{dc} \leq u_j^{dc}-b_{jh}^{dc} \leq \left(\bar{v}_j^{dc}\right)^2 \alpha_{jh}^{dc},
\end{equation}
\begin{equation}\label{eq:23f}
\left(\underline{v}_h^{dc}\right)^2 \alpha_{jh}^{dc} \leq u_h^{dc}-t_{jh}^{dc} \leq \left(\bar{v}_h^{dc}\right)^2 \alpha_{jh}^{dc},
\end{equation}
\end{subequations}
\[
\forall j, h \in \mathcal{N}^{dc}, \quad \forall (jh),(hj) \in \mathcal{E}^{dc}
\]
Where: $\mathrm{M}$ is a sufficiently large positive constant. $ b_{jh}^{dc}$ and $ t_{jh}^{dc}$ are newly introduced auxiliary variables. Similar to $\alpha_{ij}^{dc}$, both $ b_{jh}^{dc}$ and $ t_{jh}^{dc}$ exhibit symmetry properties, such that $ b_{jh}^{dc} = b_{hj}^{dc}$ and $ t_{jh}^{dc} =  t_{hj}^{dc} $. It can be deduced that when  $\alpha_{ij}^{dc}:=1$, $u_j^{d c}-u_h^{d c}=r_{j h}^{d c}\left(p_{j h}^{d c}-p_{h j}^{d c}\right)$ is activated and otherwise inactivated.

However, the TS formulation can be further refined by explicitly considering breaker actions. We know that breakers are installed at both ends of each branch for protection: \textit{One branch is disconnected if either of the two breakers is open, while it remains connected only when both breakers are closed}. We introduce binary $\beta_{jh}^{dc}$ and $\beta_{hj}^{dc}$ to represent the statuses of the breakers installed at the two ends of branch $jh$. Specifically, $\beta_{jh}^{dc}$ denotes the breaker located near bus $j$, while $\beta_{hj}^{dc}$ denotes the breaker located near bus $h$. Let $\beta_{jh}^{dc}:=1$ indicate that the corresponding breaker is closed, and $\beta_{jh}^{dc}:=0$ indicate that it is open. We have the following constraints:  
\begin{equation}\label{eq:24}
    \alpha_{jh}^{dc} \leq \beta_{jh}^{dc}, \quad \alpha_{jh}^{dc} \leq \beta_{hj}^{dc}, \quad \alpha_{jh}^{dc} \geq \beta_{jh}^{dc}+\beta_{hj}^{dc}-1.
\end{equation}
\[
 \forall (jh),(hj) \in \mathcal{E}^{dc}
 \]
Where: \eqref{eq:24} formulates the influence of breaker operations on the DC branch connection status. It can be observed that $\alpha_{jh}^{dc}=1$ holds only when both $\beta_{jh}^{dc}:=1$ and $\beta_{hj}^{dc}:=1$. Otherwise, $\alpha_{jh}^{dc}=0$.
 
\subsection{Constraints for Information Flow }
\begin{figure}
    \centering
    \includegraphics[width=0.85\linewidth]{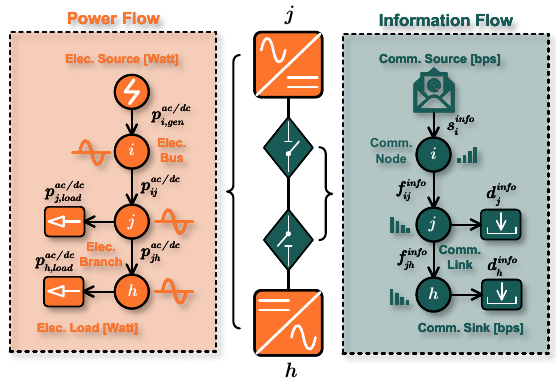}
    \caption{Analogy between power flow and information flow. From a graph-theoretic perspective, information flow is similar to power flow: communication nodes correspond to electrical buses, communication links to electrical branches, information sources to electrical sources, and information sinks to electrical loads.}
    \label{fig:2}
\end{figure}

In AC/MTDC grids, achieving optimal TS requires system-wide coordination. This implies that a breaker executes its action only after the associated communication node has received the required control signal. To this end, this work incorporate a set of basic information-flow constraints to capture, to some extent, the influence of communication on electrical system operations. As presented in Fig.~\ref{fig:2}, information flow is analogous to power flow, and the corresponding constraints are formulated as follows.

\begin{equation}\label{eq:25}
\sum_{ij} f_{ij}^{\mathit{info}} =
s_i^{\mathit{info}} - d_i^{\mathit{info}},
\end{equation}
\begin{equation}\label{eq:26}
-\bar{f}_{ij}^{\mathit{info}} \le f_{ij}^{\mathit{info}} \le \bar{f}_{ij}^{\mathit{info}},
\end{equation}
\[
\forall i \in \mathcal{N}^{\mathit{info}}, \quad \forall (ij)\in \mathcal{E}^{\mathit{info}}
\]
Where: $\mathcal{N}^{\mathit{info}}$ and $\mathcal{E}^{\mathit{info}}$ denote the sets of nodes and edges in the information network, respectively. In \eqref{eq:25}, $f_{ij}^{\mathit{info}}$ represents the information flow along the communication link. $s_i^{\mathit{info}}$ and $d_i^{\mathit{info}}$ denote the communication source and sink, respectively. A reference flow direction is predefined, such that $f_{ij}^{\mathit{info}} = -f_{ji}^{\mathit{info}}$. Therefore, the information flow on each communication link can only take one actual direction at a time. $\bar{f}_{ij}^{\mathit{info}}$ denotes the corresponding link capacity limit. Since the sign of $f_{ij}^{\mathit{info}}$ determines the actual flow direction, the information flow may take either positive or negative values. Therefore, the link capacity constraint must bound the flow magnitude in both directions, as formulated in \eqref{eq:26}.

Let us further focus on the coupling between information flow and power flow. For each breaker, the corresponding breaker-status variable $\beta_{jh}^{\mathit{info}}$ is governed by the information flow. In other words, the state of $\beta_{jh}^{\mathit{info}}$ can change only when the corresponding communication demand is satisfied, i.e., the corresponding control signal is successfully received. Such a state transition may correspond to either opening the breaker, i.e., $\beta_{jh}^{\mathit{info}}: 1 \rightarrow 0$, or closing the breaker, i.e., $\beta_{jh}^{\mathit{info}}: 0 \rightarrow 1$. Here, $(\hat{\bullet})$ is used to denote the known previous status of $\beta_{jh}^{\mathit{info}}$. Accordingly, the following formulations are introduced to describe the coupling between information flow and power flow.
\begin{subequations}\label{eq:28}
\begin{equation}\label{eq:28a}
\begin{aligned}
\gamma_{jh}^{\mathit{info}} &\ge \beta_{jh}^{dc}-\hat\beta_{jh}^{dc}, \quad
\gamma_{jh}^{\mathit{info}} \ge \hat\beta_{jh}^{dc}-\beta_{jh}^{dc},
\end{aligned}
\end{equation}
\begin{equation}\label{eq:28b}
\begin{aligned}
\gamma_{jh}^{\mathit{info}} &\le \beta_{jh}^{dc}+\hat\beta_{jh}^{dc}, \quad
\gamma_{jh}^{\mathit{info}} \le 2-\beta_{jh}^{dc}-\hat\beta_{jh}^{dc},
\end{aligned}
\end{equation}
\end{subequations}
\begin{equation}\label{eq:29}
d_j^{\mathit{info}}=\sum_{jh} d_{jh}^{\mathit{info}} \gamma_{jh}^{\mathit{info}}.
\end{equation}
\[
\forall j \in \mathcal{N}^{\mathit{info}}, \quad \forall (jh) \in \mathcal{E}^{dc}
\]
Where: Binary variables $\gamma_{jh}^{\mathit{info}}$ are introduced to represent whether the communication demand associated with breaker $jh$ is activated. Here, $d_{jh}^{\mathit{info}}$ denotes the communication demand required to control the breaker installed on branch $jh$ near bus $j$. It can be observed that a breaker action is required only when the breaker state changes, i.e., $\beta_{jh}^{dc} \neq \hat{\beta}_{jh}^{dc}$. Under this condition, \eqref{eq:28} enforces $\gamma_{jh}^{\mathit{info}} = 1$, indicating that the corresponding communication demand must be satisfied for the control signal to be delivered successfully. By contrast, when $\beta_{jh}^{dc} = \hat{\beta}_{jh}^{dc}$, the breaker remains in its original state and no communication-assisted switching action is required. In this case, \eqref{eq:28} enforces $\gamma_{jh}^{\mathit{info}} = 0$. For example, consider the case where $\beta_{jh}^{dc} \neq \hat{\beta}_{jh}^{dc}$ and assume that $\hat{\beta}_{jh}^{dc}:=1$. In this case, $\beta_{jh}^{dc}$ can only take the value  $\beta_{jh}^{dc}=0$. In this case, \eqref{eq:28} enforces $\gamma_{jh}^{\mathit{info}}=1$. Consequently, in \eqref{eq:29}, the corresponding communication demand $d_{jh}^{\mathit{info}}$ also remains activated. Similar conclusions can be verified for the other possible cases as well.

\subsection{Optimization Goals}
Based on the above, the constraints summarized in Tables~\ref{tab:2}--\ref{tab:5}, together with \eqref{eq:23}--\eqref{eq:28}, jointly constitute the complete OIPF constraints. Regarding the optimization objective, the proposed OIPF incorporates generation cost $Cost^{st}$, transmission switching cost $Cost^{nd}$, and communication cost $Cost^{rd}$. For each kind of the cost, the explicit formulations are given below. 
\begin{equation}
\begin{aligned}
Cost^{st}:=
\sum_i
&
a_{i,gen}^{ac}\left(p_{i,gen}^{ac}\right)^2
+b_{i,gen}^{ac}p_{i,gen}^{ac}
\\
&
+c_{i,gen}^{ac},
\quad
\forall i\in\mathcal{N}^{ac}
\end{aligned}
\end{equation}
Where: $a_{i, g e n}^{ac}/b_{i, g e n}^{ac}/c_{i, g e n}^{ac}$ is the quadratic/linear/constant generation cost coefficient for the generator at AC node $i$.
\begin{equation}
Cost^{n d}:=\sum_{j h} \lambda_{j h}^{d c} \varepsilon_{j h}^{d c}, \quad \forall (jh) \in \mathcal{E}^{dc}
\end{equation}
with
\begin{subequations}\label{eq:32}
\begin{align}
\varepsilon_{jh}^{dc} &\geq \beta_{jh}^{dc}-\hat{\beta}_{jh}^{dc}, \quad
\varepsilon_{jh}^{dc} \geq \hat{\beta}_{jh}^{dc}-\beta_{jh}^{dc}, \label{eq:32a}\\
\varepsilon_{jh}^{dc} &\leq \beta_{jh}^{dc}+\hat{\beta}_{jh}^{dc}, \quad
\varepsilon_{jh}^{dc} \leq 2-\beta_{jh}^{dc}-\hat{\beta}_{jh}^{dc}. \label{eq:32b}
\end{align}
\end{subequations}
Where: $\lambda_{jh}^{dc}$ denotes the line switching cost for the DC branch $jh$. $\epsilon_{jh}^{dc}$ is the binary variable. Its function is similar to $\gamma_{jh}^{dc}$, so that \eqref{eq:32} has a similar expression to \eqref{eq:28}. 
\begin{equation}\label{eq:34}
Cost^{rd}:=\sum_{jh} \lambda_{jh}^{\mathit{info}} \max\left(f_{jh}^{\mathit{info}},\,0\right), \quad \forall (jh) \in \mathcal{E}^{\mathit{info}}
\end{equation}
Where: $\lambda_{jh}^{\mathit{info}}$ denotes the communication cost associated with the communication link $jh$. The operator $\max(\cdot,0)$ is introduced to ensure that only positive information flow contributes to the total communication cost. This is necessary because $f_{jh}^{\mathit{info}}$ is allowed to take both positive and negative values to represent bidirectional information flow on the same communication link, as described in \eqref{eq:26}. Since communication cost is incurred only for the direction along which information is actually transmitted,  $\max(\cdot,0)$ ensures that only the utilized transmission direction contributes to the objective function, thereby avoiding duplicated cost accounting.

However, \eqref{eq:34} is not a standard formulation in mathematical programming. Therefore, an auxiliary variable $f_{jh}^{\mathit{info},+}$ is introduced, through which \eqref{eq:34} can be equivalently reformulated as follows:
\begin{equation}\label{eq:35}
Cost^{rd} := \sum_{jh}
\lambda_{jh}^{\mathit{info}} f_{jh}^{\mathit{info},+}, \quad \forall (jh) \in \mathcal{E}^{\mathit{info}}
\end{equation} 
with
\begin{equation}\label{eq:36}
\begin{aligned}
f_{jh}^{\mathit{info},+} \ge f_{jh}^{\mathit{info}}, \quad
f_{jh}^{\mathit{info},+} \ge 0.
\end{aligned}
\end{equation}

\section{Case Study} \label{section:3}
\begin{figure}
    \centering
    \includegraphics[width=0.9\linewidth]{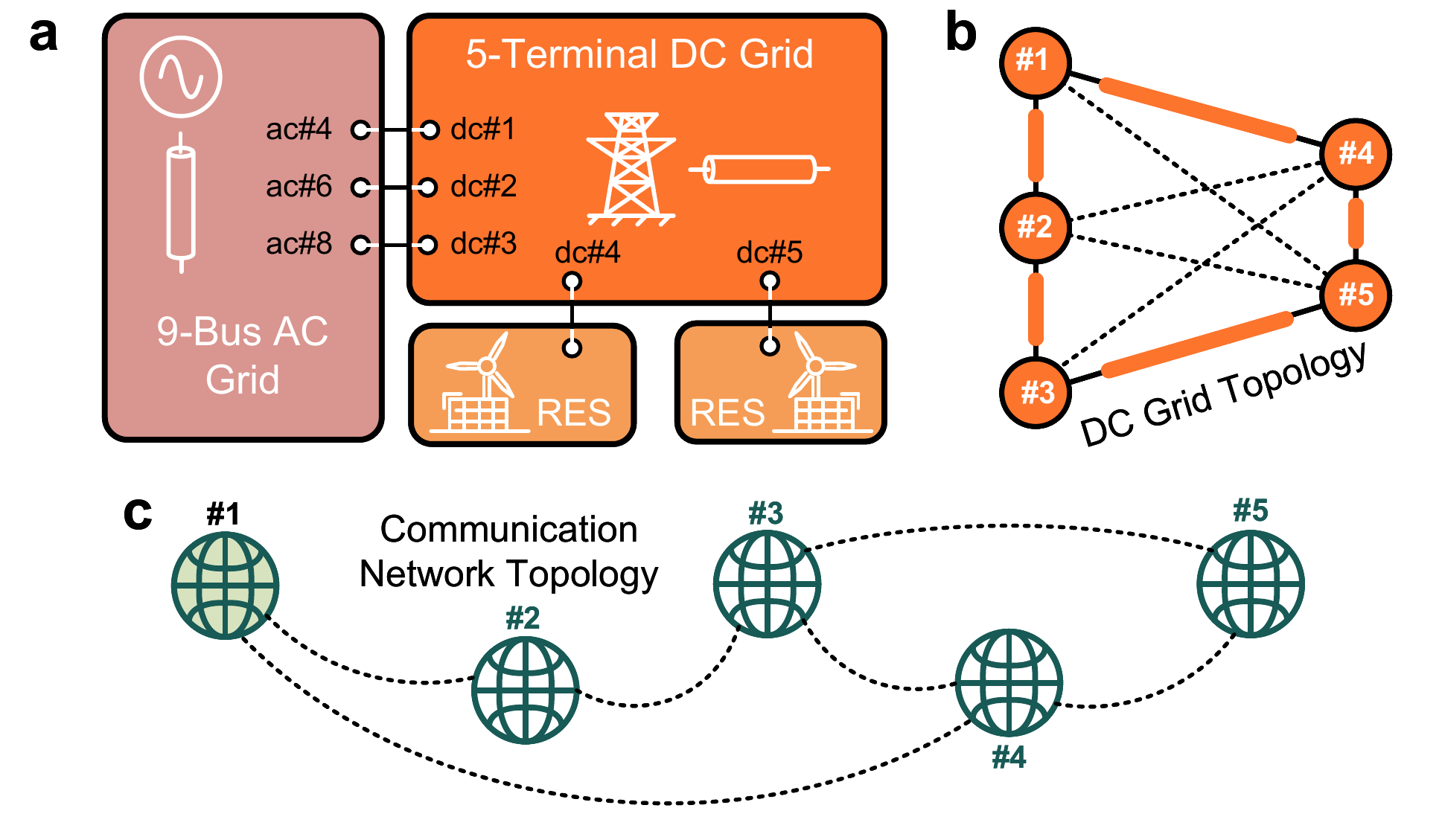}
    \caption{Case system. (a) presents the AC/MTDC grid. The MTDC grid has five terminals. Three terminals are connected to the 9-bus AC grid (Its topology is consistent with the IEEE-9 benchmark), and two terminals are connected to RES plants. (b) shows the 5-terminal DC topology. Branches \#1–\#2, \#2–\#3, \#1–\#4, \#3–\#5, and \#4–\#5 are in service. The candidate Branches \#1–\#5, \#2–\#4, \#2–\#5, and \#3–\#4 can be activated to support topology reconfiguration. (c) shows the communication network topology. Node \#1 is assumed to be the communication source, while others are communication sinks.}
    \label{fig:3}
\end{figure}

We use the case system illustrated in Fig.~\ref{fig:3} for the numerical experiments. For comparison, we have
\begin{itemize}
    \item \textit{OPF model}: The objective is set to $Cost^{s t}+Cost^{n d}$. Constraints include \eqref{eq:1}–\eqref{eq:8}, \eqref{eq:10}–\eqref{eq:24}.
    \item \textit{OIPF model}: The objective is set to $Cost^{s t}+Cost^{n d}+Cost^{r d}$. Constraints include \eqref{eq:1}–\eqref{eq:8}, \eqref{eq:10}–\eqref{eq:29}. 
\end{itemize}
Both OPF and OIPF models are typically MISOCP problems and solved by \textsc{Gurobi}\footnote{\url{https://www.gurobi.com}}. 

We first verify the rationality of the relaxed AC/DC power flow formulation. The standard nonlinear AC/DC power flow NLP model solved by \textsc{Ipopt}\footnote{\url{https://coin-or.github.io/Ipopt/}} is used as the benchmark. For a fair comparison, constraints \eqref{eq:23} and \eqref{eq:24} are excluded, and only $Cost^{st}$ is kept in the \textit{OPF model} as the optimization objective. In this way, the original MISOCP problem is reduced to an SOCP problem, ensuring that the NLP and SOCP models differ only in their AC/DC power flow formulations. As presented in Table~\ref{tab:pf}, the generation results obtained from the SOCP model are very close to those of the NLP model. Meanwhile, the SOCP model requires significantly less computational effort, making it a favorable choice for practical applications.

\begin{table}[t]
\centering
\caption{Comparison of AC/DC OPF results between NLP and SOCP Models}
\label{tab:pf}

\begin{tabular}{lcc}
\Xhline{0.75pt}
\textbf{Item} & \textbf{NLP Model} & \textbf{SOCP Model} \\
\Xhline{0.5pt}
Generation Cost [\$] & 2529.74 & 2518.07 \\
$P_{G1}$ [MW]   & 43.63 & 43.38 \\
$P_{G2}$ [MW]   & 77.09 & 76.77 \\
$P_{G3}$ [MW]   & 54.59 & 54.35 \\
Solving Time [s] & 0.78 & 0.02 \\
\Xhline{0.75pt}
\end{tabular}
\vspace{0.5pt}
\begin{flushleft}
\footnotesize
\ding{71} A warm-up run was first conducted to reduce initialization effects, and the reported computation time is averaged over five subsequent runs. Tests were conducted on a laptop with an Intel Core i9-12900HK Processor.
\end{flushleft}
\end{table}

\begin{figure*}
    \centering
    \includegraphics[width=\textwidth]{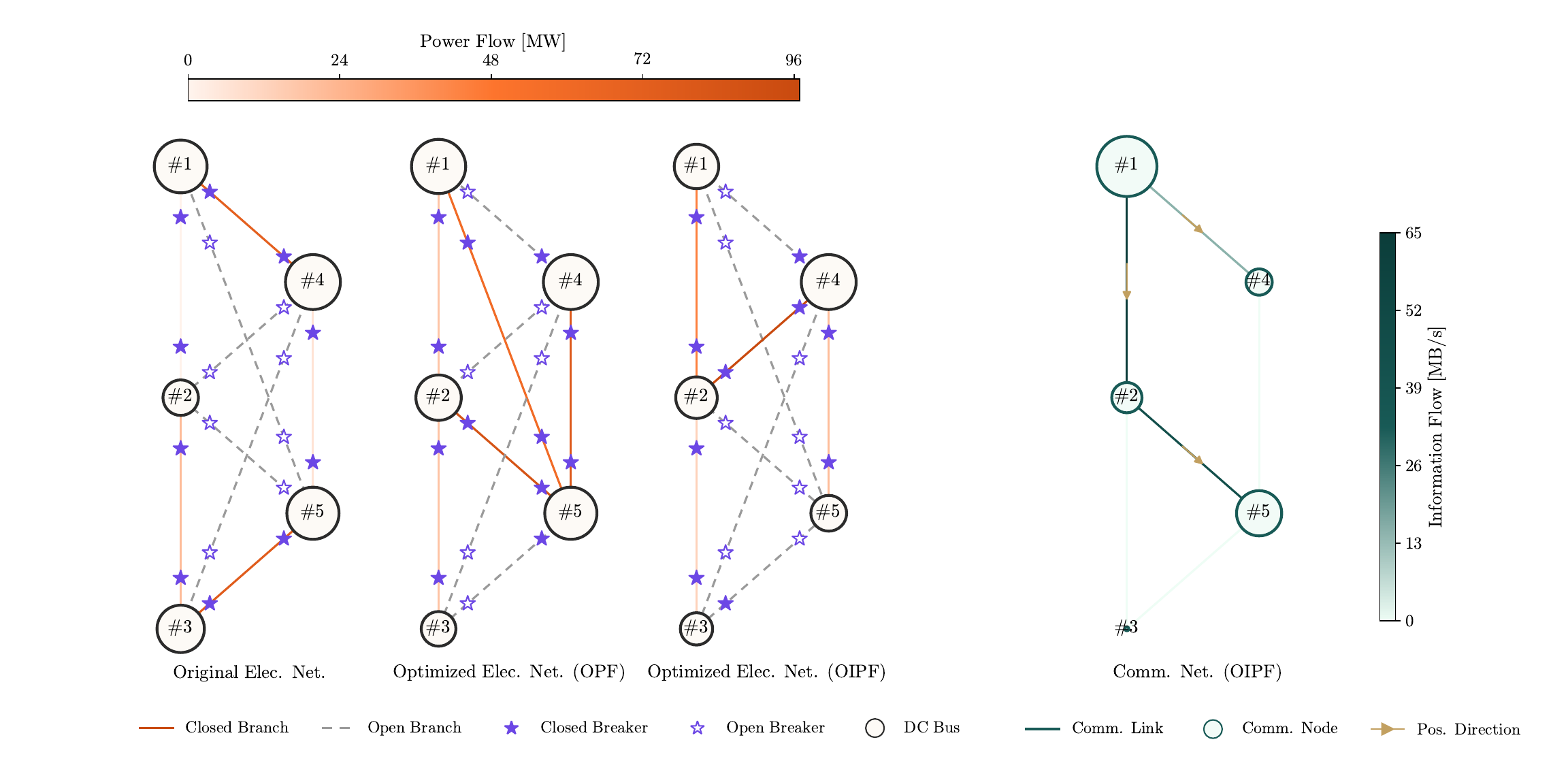}
    \caption{TS results. For both electrical and communication networks, darker (lighter) branch colors indicate larger (smaller) flow magnitudes, and larger (smaller) node sizes represent higher (lower) injection levels. Note that the branch power flow is plotted according to the averaging that $(|p_{jh}^{dc}| + |p_{hj}^{dc}|)/2$. }
    \label{fig:4}
\end{figure*}

We further investigate the impact of information flow. Table~\ref{tab:6} summarizes the key parameters used in the numerical experiments. A scenario is considered where Branches~\#1--\#4 and \#3--\#5 are faulted, interrupting RES power delivery and requiring TS for DC topology reconfiguration. As shown in Fig.~\ref{fig:4}, OPF activates Branches~\#1--\#5 and \#2--\#5, while additional branches are not selected since their switching costs cannot justify the reduction in generation cost. OPF also tends to operate breakers near Bus~\#1 and Bus~\#3, whose switching costs are lower according to Table~\ref{tab:6}. By contrast, OIPF selects only Branch~\#2--\#4, since activating Branches~\#1--\#5 and \#2--\#5 incurs higher communication costs ($60+55=115$ and $70+65=115$~MB/s), whereas Branch~\#2--\#4 requires only $20+15=35$~MB/s. Consequently, RES power at Bus~\#5 is curtailed in OIPF due to the limited transmission capacity of a single activated branch. For information flow, Sink~\#3 has no communication requirement since breakers near Bus~\#3 are not operated, while Links~\#4--\#5, \#2--\#3, and \#3--\#5 carry no information flow because direct transmission through Link~\#2--\#5 is more economical. These results suggest that considering information flow is important for practical TS applications.

\begin{table*}
\centering
\caption{Lists of Key Parameters}
\label{tab:6}
\footnotesize
\begin{tabular}{ccccc}
\toprule
\textbf{From Bus} & \textbf{To Bus} & 
\makecell{\textbf{Switching Cost $\lambda_{j h}^{d c}$ [\$]} \\ (Pos. Direction / Neg. Direction)} & 
\makecell{\textbf{Breaker Comm. Demand $d_{j h}^{i n f o}$ [MB/s]} \\ (Pos. Direction / Neg. Direction)} &
\makecell{\textbf{Commun. Cost $\lambda_{j h}^{i n f o}$ [-]} \\ (Pos. Direction / Neg. Direction)} \\
\midrule
\#1 & \#2 & 150 / 100 &  20 / 15 & 15 / 15 \\
\#1 & \#4 & 250 / 300 &  40 / 35 & 25 / 30 \\
\#1 & \#5 & 100 / 100 &  60 / 55 & - / - \\
\#2 & \#3 & 200 / 300 &  30 / 25 & 20 / 20 \\
\#2 & \#4 & 350 / 250 &  20 / 15 & - / - \\
\#2 & \#5 & 150 / 100 &  70 / 65 & 15 / 10 \\
\#3 & \#4 & 350 / 450 &  20 / 15 & - / - \\
\#3 & \#5 & 250 / 350 &  50 / 45 & 25 / 30 \\
\#4 & \#5 & 450 / 400 &  30 / 25 & 40 / 40  \\
\bottomrule
\end{tabular}
\vspace{0.5pt}
\begin{flushleft}
\footnotesize
\ding{71} ``Pos. Direction'' refers to the direction from the From Bus to the To Bus. ``Neg. Direction'' refers to the direction from the To Bus to the From Bus.
\end{flushleft}
\end{table*}

\section{Conclusion} \label{section:4}
This work proposes an MISOCP-based OIPF model for optimal TS in AC/MTDC grids. Case studies reveal that considering communication costs can lead to TS decisions that differ significantly from those based solely on OPF. Although the communication network modeling remains preliminary, future work will incorporate communication delays to improve model realism and support more effective TS decisions.

\bibliographystyle{IEEEtran}
\bibliography{ref}

\end{document}